# TOWARD A HUMAN-CENTERED AI-ASSISTED COLONOSCOPY SYSTEM

**Hsiang-Ting Chen, Yuan Zhang, Gustavo Carneiro**
University of Adelaide
Adelaide, Australia
{tim.chen, yuan.zhang01, gustavo.carneiro}@adelaide.edu.au

**Seon Ho Shin, Rajvinder Singh**
Lyell McEwin Hospital
Adelaide, Australia
{rajvinder.singh,seonho.shin}@sa.gov.au

## ABSTRACT

AI-assisted colonoscopy has received lots of attention in the last decade. Several randomised clinical trials in the previous two years showed exciting results of the improving detection rate of polyps. However, current commercial AI-assisted colonoscopy systems focus on providing visual assistance for detecting polyps during colonoscopy. There is a lack of understanding of the needs of gastroenterologists and the usability issues of these systems. This paper aims to introduce the recent development and deployment of commercial AI-assisted colonoscopy systems to the HCI community, identify gaps between the expectation of the clinicians and the capabilities of the commercial systems, and highlight some unique challenges in Australia.

## 1 Introduction

Bowel cancer is the second most common and deadliest cancer in Australia [1]. The National Bowel Cancer Screening Program (NBCSP) reduces the cancer burden by recommending those with high risk criteria or a positive faecal occult blood test (FOBT) to undergo a surveillance colonoscopy. The main purpose of the colonoscopy procedure is to find and remove abnormal growth (i.e. premalignant adenomatous polyps) before it turns into bowel cancer [2]. However, colonoscopy is not a perfect test, as polyps can often be missed or misdiagnosed as a result of human error. The chance of detecting a polyp depends on the endoscopist's technique, experience, bowel preparation (cleanliness), and concentration/fatigue. All endoscopists aim to achieve a high adenoma detection rate (ADR), as this is a proven performance indicator and has an inverse relationship with cancer risk [3]. When a polyp is discovered, an endoscopist must visually inspect and classify (diagnose) the polyp as either benign, adenomatous or cancerous, and choose the polyp management accordingly. The classification of polyps is an expert-based decision based on the surface appearance of the polyp. However, in challenging cases, endoscopists cannot make a reliable optical diagnosis and resort to a histopathological diagnosis of biopsied or resected polyp. This routine practice carries potential harm to patients and high costs to the health system. Recent meta-analysis reports a high rate of missed polyps (between 22% and 27% [4, 5]) and of missed bowel cancers, where up to 8% are not detected during colonoscopy [6].

In parallel, AI models for polyp detection and classification have advanced significantly in the last decade. The machine- and deep-learning-based models have shown surprisingly impressive results in both retrospective and prospective setups [7]. Recently, several randomised controlled trials also affirmed the effectiveness of CADe systems. The meta-analysis suggests a 50% increase in ADR, which will significantly benefit cancer detection. These outcomes have given major colonoscopy device manufacturers enough confidence to accelerate the deployment of commercial AI products that provide visual guides in real-time to assist the polyp detection and classification.

Despite promising results from early clinical trials, these products raise questions about integrating AI decision support into daily clinical practice. Endoscopists, who are early adopters of these AI products, have expressed AI fatigue caused by too much information provided by the AI system [8]. They have also shown concern that the lack of transparency of the AI model might lead to incorrect or missing recommendations that can cause patient harm.

This paper aims to introduce the development and deployment of AI-assisted colonoscopy systems to the HCI community, identify gaps between the expectation of the clinicians and the capabilities of the commercial systems, and highlight some unique challenges in Australia. To this end, we conducted a field study at the [hospital, state] on [date]. The field study consists of observing one complete colonoscopy procedure inside the operation theatre and two semi-structured interviews with gastroenterologists with different levels of experience.The AI-assisted colonoscopy system is uniquely positioned: the AI algorithms are robust enough to obtain regulatory approvals for clinical uses and assist clinicians in making decisions with significant consequences; yet, the usability of the AI system is considered poor by many clinicians [8]. We believe the AI-assisted colonoscopy field direly needs the HCI community's attention. The introduction of a human-centred design approach can significantly increase clinicians' adoption and acceptance and lead to better outcomes for patients in Australia and around the world.

## 2 Background and Related Work

This section aims to provide a brief overview of AI-assisted colonoscopy to the HCI community. The section will cover the colonoscopy procedure, state-of-the-art machine learning methods for polyp detection, commercial AI-assisted colonoscopy systems, and the recent efforts on AI-assisted colonoscopy from the HCI community.

### 2.1 Colonoscopy

Colorectal cancer (CRC) is a common malignant tumour of the digestive tract that develops in the colon. According to the Cancer in Australia 2019 [9], colon cancer was newly diagnosed in 16,398 individuals and caused 5,597 deaths, which was the deadliest and second most commonly diagnosed cancer in Australia. CRC mostly begins as **polyps**, which can be flat, raised or on a stalk of variable length. The polyps can also present as carpet-like abnormal growth along the colonic lumen. According to the International Colorectal Endoscopic (NICE) Classification, polyps can be broadly categorised into three types:

- **Hyperplastic polyps (type 1)** form about 20% to 40% of all polyps. These polyps are generally considered benign polyps and carry a very low risk of malignant change.
- **Adenomatous polyps (type 2)** form about 70% of all polyps and are regarded as premalignant. Most of these polyps do not become cancerous but has the propensity to develop into colorectal cancer. The risk of containing a focus of cancer increases significantly for polyps with a diameter exceeding 10mm. A subgroup of adenomatous polyps is *sessile serrated adenomas*, which can progress to cancer rapidly and are difficult to detect due to their flat shape and subtle features with an overlying mucous layer.
- **Malignant polyps (type 3)**, also called cancerous polyps, contains cancerous cells and forms colorectal cancer.

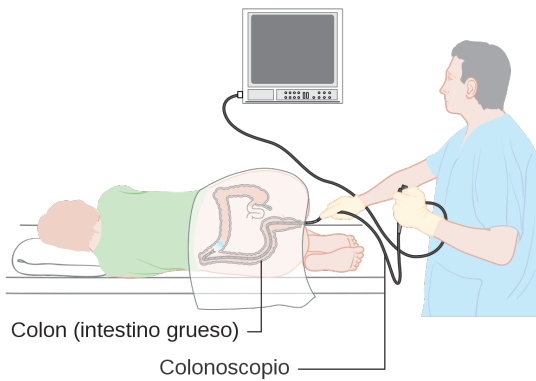

Figure 1: The colonoscopy procedure diagram.

Colonoscopy is the most effective way to reduce the incidence and mortality of colorectal cancer [2]. The colonoscopy usually lasts 30 to 60 minutes and examines the colon with a fibre optic camera on a flexible tube passed through the anus, as depicted in Figure 1. The videos from the camera are displayed on an external screen and allow endoscopists to conduct a visual diagnosis. When a polyp is discovered, the endoscopist visually inspects and diagnoses the polyp as either hyperplastic, adenomatous or malignant and decides the corresponding management of these polyps. In the case where a reliable optical diagnosis cannot be reached, the endoscopist would resort to a histopathological assessment of polyps by biopsying or retrieving resected polyps for diagnosis. This routine practice carries potential harm to patients, inefficiencies and high costs.

Colorectal cancer is prevented if the adenomatous polyp is found and removed before turning malignant. The National Bowel Cancer Screening Program reduces the cancer burden by recommending colonoscopy screening in patients with high-risk criteria or who have a positive faecal occult blood test. However, despite increasing numbers of colonoscopies performed per capita, there is a persistent heavy burden of colorectal cancer [10]. Previous studies showed that the missing rates for total polyps and adenomas in colonoscopy were 16.8-28% and 16-26%, respectively. The missing rate for advanced adenomas might reach as high as 11%, and these missing adenomas are considered to be the direct cause of 50-60% of interval cancer [11].

The chance of correct detection and classification of polyps during a colonoscopy is dependent on the endoscopist (experience, concentration, fatigue), the operation-related factors (bowel preparation, scope withdrawal time), and the characteristics of the lesion [12]. In the last decades, researchers have proposed new endoscopic hardware to improve polyp detection during colonoscopies, such as high-definition colonoscopy, narrow-band imaging (NBI) technology and intelligent colour enhancement technology [13, 14, 15]. Although these new endoscopic technologies have improved the visual field of colonoscopy and the recognition of mucosal lesions to some extent, their role in improving polyp detection is still controversial because these technologies require more experienced gastroenterologists and might result in longer procedures and higher costs.

## 2.2 Computer-Aided Detection of Polyps

In the early 2000s, researchers began exploring using the computer as a *third eye* during the colonoscopy procedure to analyse the colonoscopy video in parallel with the endoscopist. Pioneering works [16, 17, 18] started applying traditional computer vision techniques such as texture analysis [16] and wavelet transformation [17] on a limited amount of static images containing polyps. In 2012, Bernal et al. [19] presented one of the first successful applications of computer-assisted detection (CADe) of polyps using classical computer vision techniques (non-learning-based methods) and achieved an AUC of 0.98, albeit on a small data set with non-interactive speed (19 seconds per frame). Since then, deep learning-based approaches have dominated computer-aided polyp detection and shown superior performance in many public competitions, such as the MICCAI 2015 challenge on automatic polyp detection [20].

We encourage readers interested in CADe to read the recent systematic review paper [7]. This paragraph only briefly describes several seminal works to demonstrate the data set's accuracy and size for the latest deep learning-based CADe works. In 2016, Li et al. presented one of the first works using the convolutional neural network (CNN) for polyp detection. The model was trained with 32,305 colonoscopy images and achieved 86% precision and 73% sensitivity [21]. Wang et al. [22] later used the SegNet architecture, trained on 5,545 images, and prospectively verified the model on 27,461 colonoscopy images. The AUC achieved by the algorithm on flat polyps (Yamada type I), smaller polyps (<=0.5cm) and isochromatic polyps is 0.943, 0.957 and 0.957, respectively. In 2019, Yamada et al. [23] presented an impressive system that detects polyp with 97.3% sensitivity and 99.0% specificity at a real-time speed of 45 frames per second. Later works [24, 25, 26] expanded the training data to 100,000 images and achieved 95% sensitivity and specificity.

## 2.3 Commercial AI-assisted Colonoscopy Systems

The impressive results of recent AI models have led to several commercial AI-assisted colonoscopy systems in the market. Notably the *GI-Genius* by Medtronic [1], the *EndoBRAIN* by Cybernet [2], *AI4GI* by Satisfai [3], and the *CAD EYE* from Fujifilm [4]. All these AI-assisted colonoscopy systems are distributed as extension modules for the existing endoscope system from either Olympus or Fujifilm. The AI module takes the video stream from the endoscope as an input and performs polyp detection using CNN-based algorithms and polyp tracking across frames utilising some variation of the YOLO algorithms [27] in real-time.

The CADe assistance comes in various forms (Figure 2). All modules provide the capability of directly rendering the CADe information over the video stream that is linked to the endoscope display. The GI-Genius and AI4GI modules use the bounding box visualisation, which is common in object tracking applications. Once a polyp is detected and localised, a bounding box is drawn at the location of the polyp (Figure 2a). The EndoBRAIN made the design decision to avoid obstructing the endoscopists' view of the colons by keeping the endoscope video stream intact. Once a polyp is detected, the module evokes an alert sound and highlights the four corners of the colonoscopy video in yellow colour (Figure 2b). The CAD EYE system provides a suite of display options. The module provides the bounding boxes, similar to GI-Genius and AI4GI module, the visual assistance circle that surrounds the colonoscopy stream and lights up when a polyp is detected, similar to EndoBrain EYE module, and a position map that indicate the positions of the suspected polyps in a separate sub-window (bottom-right in Figure 2d).

---

[1] https://www.medtronic.com/covidien/en-us/products/gastrointestinal-artificial-intelligence/gi-genius-intelligent-endoscopy.html
[2] https://www.cybernet.co.jp/medical-imaging/products/endobrain/
[3] https://satisfai.health/gi-cancer-treatments
[4] https://fujifilm-endoscopy.com/cadeye

EndoBRAIN and CAD EYE systems also provide extra modules which link to the polyp detection module and perform an additional classification step on the detected polyps in real-time. The *EndoBRAIN-plus* module classifies the detected polyp into non-neoplastic (hyperplastic), adenoma, or invasive cancer (malignant). The *CAD EYE Characterisation* module provides a binary polyp classification of either hyperplastic (benign) or neoplastic (malignant). Both modules display the classification results beside the colonoscopy video stream in text. For example, the white text under Figure 2d suggests that the AI classifies the polyps on display as neoplastic (malignant).

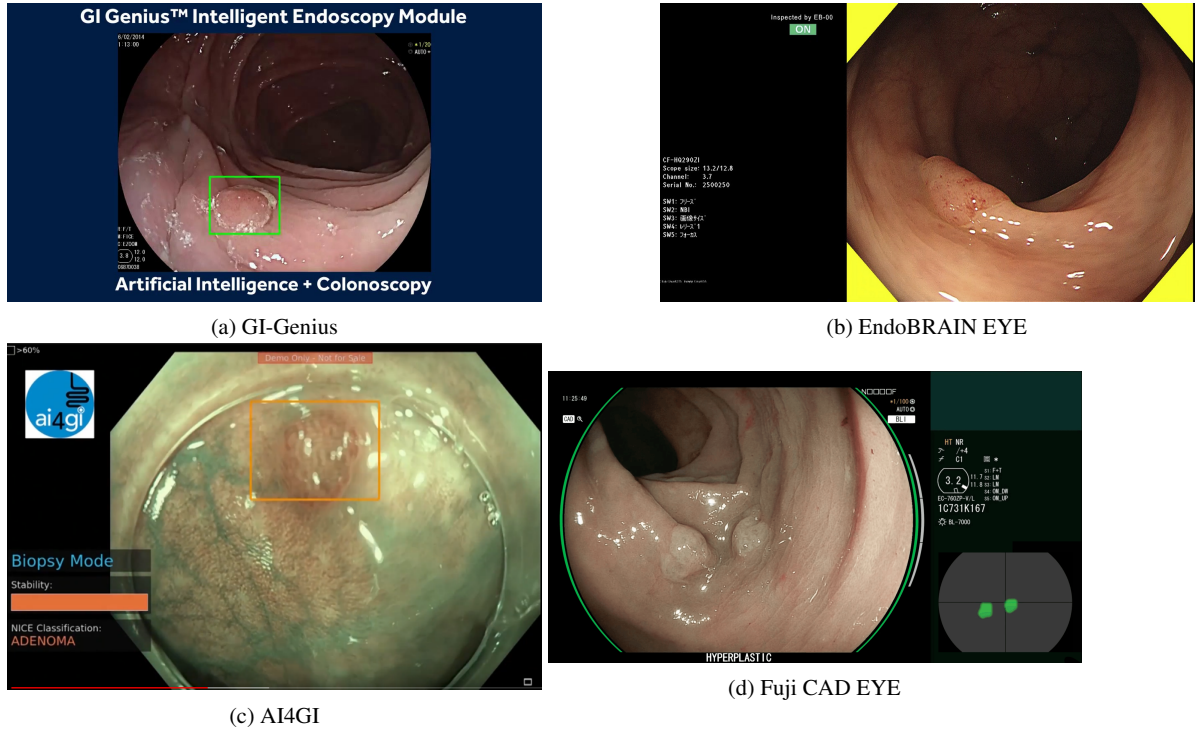

(a) GI-Genius

(b) EndoBRAIN EYE

(c) AI4GI

(d) Fuji CAD EYE

Figure 2: Commercial AI-assisted colonoscopy systems

Based on these commercial systems, there were five randomised controlled trials published so far [7]. These trials showed that the group using CADe had a higher ADR (33.6%) than the control group (25.2%). The lesion detection is also not affected by factors such as size. Current evidence suggests clear benefits of integrating AI into the colonoscopy procedure.

### 2.4 Human-AI Interaction

The interaction between humans and AIs has long been a core research topic in the HCI community. As the machine learning community made the paradigm shift into deep learning-based models in the last decade, new interaction challenges have emerged [28]. The traditional machine learning pipeline usually starts with handcrafting a list of features or rules, which are then fed into machine learning models, such as regression or decision trees. The manual selection of features or rules depends heavily on domain knowledge of the question and could be a tedious trial-and-error process. Still, these manually engineered features and rules will likely be understandable to humans. In contrast, the modern deep learning-based pipeline employs an end-to-end approach, which skips the feature engineering step and relies on a massive amount of labelled data to fine-tune the parameters of neural networks. Such an end-to-end approach results in overall better performance. However, the neural network itself does not provide human-understandable semantics, and its high dimensionality often renders traditional visualisation methods less effective [29]. How to interact with *black box* AIs that outperform humans in specific tasks rapidly becomes a topic of interest to HCI researchers. Indeed, the HCI community has made significant progress in framing the human-AI interaction problem in the last five years. The conceptualisation of the human-centred AI framework [30], and the establishment of human-AI interaction guidelines from practical engineering experience [31] are perhaps the two most prominent achievements, among many others.

Among all human-AI interaction scenarios, AI in medical health has long been the point of contention [32]. The current consensus of practitioners and AI researchers could be nicely summarised by the famous quote: "AI won't replace radiologists, but radiologists who use AI will replace radiologists who don't." [33]. Toward augmenting the clinicians, the HCI community has applied human-centred AI methodology to many different medical practices. For examples: CheXplain [34] assists physicians in exploring and understanding chest X-ray analysis. Unremarkable AI [35] argues for an unobtrusive presentation even for AI prediction with significant consequences. Cai et al. [36] introduced a novel *refine-by-concept* tool, which expands the traditional medical image search beyond visual similarity. Wang et al. [37] proposed a theory-driven conceptual framework for explainable AI and consolidated the design by conducting a co-design process with the clinician on ICU data.

Few works investigated colonoscopy using HCI methodologies. Van Berkel et al. [38] evaluated seven visual markers for polyp detection and concluded that the clinicians preferred blue wide bounding circles. Notably, the experiment did not assess the position map design from the CAD EYE system (Figure 2d). AI-assisted colonoscopy has also been discussed as a case study under the theme of continuous human-AI interaction [39, 40].

## 3 Field Study

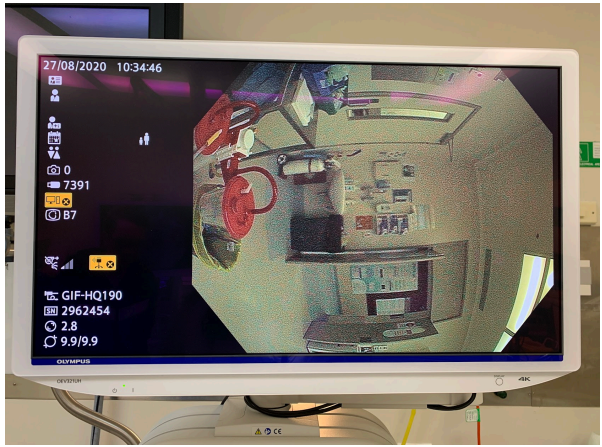

Figure 3: The endoscope display at the [hospital]

We conducted a field study to understand the colonoscopy procedure and how an AI system might assist endoscopists in Australia. The field study took place at the [anonymised hospital, state] on [anonymised date] by a team consisting of one user experience designer and two AI researchers. The field study started with an observation of a colonoscopy procedure inside an operation theatre. One senior gastroenterologist performed the colonoscopy procedure with a junior gastroenterologist, a nurse and an anaesthetist. The endoscope equipment in the operation room was an Olympus X1 system with a 4K display (Figure 3). The observed procedure did not involve the use of an AI-assisted colonoscopy system, which is still rare in Australia. The procedure lasted about 30 minutes. After the observation, we conducted two separate semi-structured interviews with one senior and one junior gastroenterologist. Both gastroenterologists are specialist registrars and have knowledge of the CADe system. The senior gastroenterologist had experience using a CAD EYE system from Fujifilm. The interview questions that guide the conversation can be found in the Appendix ??.

## 4 Result and Discussion

In this section, we report our findings of the field study. We first describe the observed colonoscopy procedure and then summarise the interviewee's responses to different questions. We will also discuss the endoscopists' expectations and concerns about AI-assisted colonoscopy and the unique challenges of the deployment of AI-assisted colonoscopy in Australia.

### 4.1 Colonoscopy Procedure

We summarise the colonoscopy procedure learned from the direct observation and the interview in Figure 4. The colonoscopy procedure begins with the endoscopist inserting the endoscope into the colon. Through the visual inspection, the endoscopist needs to continue the intubation until the endoscope reaches the caecum,i.e. the tip of the endoscope touches the appendiceal orifice. After reaching the caecum, the endoscopist slowly withdraws the endoscope and examines the colon's inner lining. When the endoscopist found a polyp, the endoscopist would maneuver the endoscope for a close-up view of the polyp and takes a screenshot for documentation. The endoscopist then classifies the polyp based on the visual inspection and decides a treatment, such as no intervention or removal and retrieval for pathology analysis. The polyp detection, diagnosis, and treatment process continues until the endoscope is wholly withdrawn. After the procedure is completed, the endoscopists then sit in front of a computer and compile a comprehensive colonoscopy report based on the captured photos during the colonoscopy. The report includes essential information such as assessments of patient risk and comorbidity, the findings of polyps and interventions, and the follow-up plan.

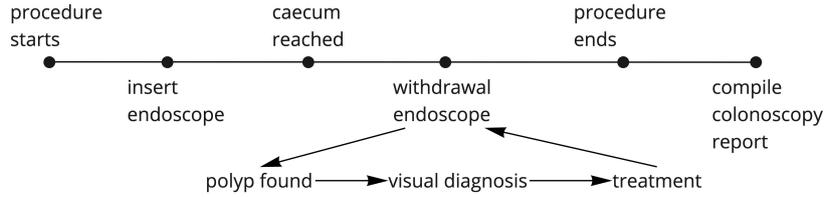

Figure 4: The colonoscopy procedure

### 4.2 Polyp Detection and Classification (Q1, Q2, Q3, Q6)

From the perspective of gastroenterologists, the shape, location and type of polyps are the main factors that affect whether polyps can be detected. Small polyps, flat polyps, or sessile serrated adenoma are challenging to spot. Moreover, polyps that grow behind folds or blind areas, including a medical wall of ascending colon, hepatic flexure, the area between the appendiceal orifice and ileocecal valve and rectosigmoid junction, are also considered difficult to detect. The distance between the polyp and the endoscope is another factor that affects the identification of polyps. Polyps that are far away from the location of the endoscope might be too little on the screen to be found. The polyps that are displayed at the peripheral of the view are also less likely to be identified.

During the interview, we also established that a difficult case for endoscopists might be a simple one for AI and vice versa. For example, a tiny polyp on the external display might be difficult for an endoscopist to detect. Yet, given the 4k resolution of the modern colonoscopy system, the polyp might be very clear to the AI model. In contrast, an AI model might be confused by different lighting or other factors such as faeces or waters in the colon. But a trained endoscopist would have a better tolerance for those conditions. Future studies will have to investigate further the feasibility of context-aware AI assistance during the colonoscopy procedure.

The interviewee also believed the combination of polyp detection and classification capability is essential to the success of AI-assisted colonoscopy. Endoscopic societies have advocated for >90% accuracy in optical diagnosis to allow cost-effective strategies (e.g., small, benign hyperplastic polyps to be ignored and adenomatous polyps to be discarded after resection, without histopathology expenses) [41]. General endoscopists have not been able to achieve this and have to remove all polyps that are then sent for a histopathological diagnosis. An AI system that can extract multiple characteristic features of polyps to make expert-level diagnostic opinions will save valuable medical resources.

### 4.3 Quality Assurance (Q4, Q5, Q7)

Beyond the detection and classification of polyps, the interviewee suggested using AIs as an auditor for the colonoscopy procedure. Poor colonoscopy withdrawal techniques, a short duration of withdrawal, poor bowel preparation, and omission to intubate the caecum all contribute to the potential misses of polyps regardless of AI assistance. For example, the optimal withdrawal time should be more than 7 minutes. An AI model might be able to recognise anatomical landmarks of the caecum and time the withdrawal process from the caecum. The AI system might also collect contextual information, such as the total number of identified polyps, and the time it takes to remove each polyp and polyp locations in the colon. The data above provide rich information to provide meaningful feedback to the endoscopist and can be used to ensure that endoscopists adhere to adequate standards.

Interviewees also expressed interest in seeing the AI system performing routine tasks such as automatically capturing screenshots of the polyps at different distances and angles as the scope approaches the polyp. The automatic estimation of polyp size, which correlates with the probability of it becoming cancerous, also has practical values. Ultimately, the interviewees would like to see a system that can automatically generate the colonoscopy procedure reports that encapsulate all the information above and summarise the procedure using natural language. Note that there are existing works on extracting medical information from manual reports [42] and the automatic generation of structured pathology reports [43]. However, few works focus on the generation of the procedure report.

**4.4 More Challenges in AI-Assisted Colonoscopy**

Recent randomised controlled trials suggest that the group using AI-assisted colonoscopy can achieve an ADR of 36.6% (vs 25.2% without AI). It has been shown that a 4% increase in ADR will reduce 1% of colorectal cancer rate during follow-up examinations [3]. This trend suggests that the use of AI might eventually be preferred by patients or even be required by law to ensure optimal outcomes. Current legal frameworks still hold the domain experts accountable for the recommendation they provide, the results they achieve, and the deliverable of quality and safety [44]. In such a context, the AI-assisted colonoscopy system should be responsible for not only helping endoscopists to detect and classify polyps. It must also assist endoscopists in fulfilling the obligation to explain, justify, and take responsibility for their final procedural decision.

On the flip side, the over-reliance on the AI system could also lead to undesired outcomes. Endoscopists may relax their rigorous search for polyps and rely on AI to detect polyps. The superior AI performance might lead to the subtle loss of self-confidence of junior doctors and lead to a bias toward AI suggestions, eventually resulting in the deskilling of junior endoscopists [45]. Future studies will have to continue to explore the effect of prolonged AI assistance and look at ways to properly introduce the AI-assisted colonoscopy system to young doctors during their training.

Australia also faces unique challenges because of the significant difference in resources and demographics between urban and rural areas. For example, most of the training data for the AI models are collected inside large and well-funded teaching hospitals, which are often located in urban areas with dense populations. It is unclear how the bias of data could affect the accuracy of the polyp detection and classification model. Similarly, many hospitals in rural areas use older colonoscopy equipment with lower-resolution images and poor image quality. Such colonoscopy images might exhibit a different distribution of features and invalidate the trained AI models. The HCI community holds interest in studying healthcare information technologies in developing countries, such as [46, 47]. We also see recent studies on the use of AI for decision support in the rural clinical context [48]. Similar efforts are needed in Australia as each region and state has unique geographic, demographic, and healthcare infrastructure. Addressing these biases in data and demographic and creating a responsible and equitable AI will be crucial to ensure Australians enjoy the benefits of recent advances in AI-enabled colonoscopy.